\documentclass[11pt]{article}

\usepackage{a4}
\usepackage{amssymb,latexsym}
\usepackage[mathscr]{eucal}
\usepackage{url}
\usepackage{ntheorem}

\newcommand{\tg}{\tilde{g}}%
\newcommand{\tnabla}{\tilde{\nabla}}%
\newcommand{\tp}{\tilde{\phi}}%
\newcommand{\tpsi}{\tilde{\psi}}%
\newcommand{\tsigma}{\tilde{\sigma}}%
\newcommand{\tR}{\widetilde{R}}%
\newcommand{\tT}{\widetilde{T}}%
\newcommand{\tm}{\tilde{m}}%

\newcommand{\tS}{\widetilde{S}}

\newcommand{\scri}{{\mathcal{I}}}%
\newcommand{\bea}{\begin{eqnarray}}

\newcommand{\eea}{\end{eqnarray}}
\newcommand{\bean}{\begin{eqnarray*}}

\newcommand{\eean}{\end{eqnarray*}}
\newcommand{\be}{\begin{equation}}

\newcommand{\ee}{\end{equation}}

\begin{document}

\title{A comment on positive mass for scalar field sources}
\author{Paul Tod\thanks{{ E--mail}: \protect\url{paul.tod@st-johns.oxford.ac.uk}}
\\
Mathematical Institute and St John's College\\ Oxford}

\maketitle


\begin{abstract}
We use a transformation due to Bekenstein to relate the ADM and Bondi masses of asymptotically-flat solutions of the Einstein equations with, respectively, scalar sources and conformal-scalar sources. 
Although the conformal-scalar energy-momentum tensor does not satisfy the Dominant Energy Condition one may, by this means, still conclude that the ADM mass is positive.
\end{abstract}

\section{Introduction}
In a recent article, \cite{BST}, the Einstein-conformal-scalar field equations were considered. These equations are known to be well-posed, and asymptotically-flat solutions exist \cite{hub}, 
but the energy-momentum tensor does not satisfy the Dominant Energy Condition (DEC), which is a hypothesis in all Positive Mass Theorems. Nonetheless, in certain cases considered in \cite{BST} 
one could still establish positivity of ADM mass and there was some evidence for positivity of Bondi mass and negativity, at least on the average, of Bondi mass-loss. In this article, we use a transformation 
which goes back at least to Bekenstein, \cite{bek1}, \cite{bek2}, (for a more recent discussion see \cite{fgn}) relating solutions of the Einstein-conformal-scalar and Einstein-scalar systems, to look at this question. Since the scalar-field energy-momentum 
tensor does satisfy DEC, we have positivity of the ADM and Bondi masses, assuming the other hypotheses of the Positive Mass Theorems, and can seek to carry 
these over to the case of the conformal scalar. In brief, we find that this works for the ADM mass, which is the same in the two cases, 
but not for the Bondi mass, which differs.

\section{The Transformation}
We take the Einstein-conformal-scalar field equations to be
\be\label{ecs1}
G_{ab}=-Q_{ab}-T_{ab}^{\mathrm{ex}},
\ee
where
\be\label{ecs2}
Q_{ab}=4\phi_{,a}\phi_{,b}-g_{ab}g^{cd}\phi_{,c}\phi_{,d}-2\phi\nabla_a\phi_{,b}-\phi^2(R_{ab}-\frac16Rg_{ab}),
\ee
which is the energy-momentum tensor for the conformal-scalar field $\phi$ which satisfies
\be\label{ecs3}
\Box\phi+\frac16R\phi=0,
\ee
(see e.g. \cite{PR} for this $Q_{ab}$) and $T_{ab}^{\mathrm{ex}}$ is any extra matter contribution which, for the transformation to work, we require to be trace-free and divergence-free. 
(Conventionally we put $8\pi G=1$ and use the sign of the Ricci tensor in \cite{PR}.)
We'll assume that $T_{ab}^{\mathrm{ex}}$ satisfies the DEC, but note that $Q_{ab}$ does not. 
By virtue of (\ref{ecs3}), $Q_{ab}$ is also trace-free and 
divergence-free.

From the trace of (\ref{ecs1}) we learn that $R=0$ and then (\ref{ecs1}) can be rearranged to become
\be\label{ecs4}
(1-\phi^2)R_{ab}=-4\phi_{,a}\phi_{,b}+g_{ab}g^{cd}\phi_{,c}\phi_{,d}+2\phi\nabla_a\phi_{,b}-T_{ab}^{\mathrm{ex}}.
\ee
Evidently, these equations are singular when $\phi^2=1$, but there do exist asymptotically-flat solutions with $\phi^2<1$ (\cite{hub}) and we shall suppose that this condition holds. The transformation 
of Bekenstein \cite{bek1}, \cite{bek2}, is found by supposing that a conformally related metric 
\[\tg_{ab}=\Theta^2g_{ab},\]
with $\Theta$ a function of $\phi$, is a solution of the Einstein-scalar field equations with scalar $\tp$ a function of $\phi$ i.e.
\be\label{es1}
\tR_{ab}=-k\tp_{,a}\tp_{,b}-\tT_{ab}^{\mathrm{ex}},
\ee
where $k$ is a number fixed by convention, which we'll take for ease to be 4, and 
\[\tT_{ab}^{\mathrm{ex}}=\Theta^{-2}T_{ab}^{\mathrm{ex}}\]
to ensure that $\tT_{ab}^{\mathrm{ex}}$ is divergence-free for $\tnabla$ (note it also satisfies the DEC). The formula for rescaling the Ricci tensor is
\[\tR_{ab}=R_{ab}+2\Theta^{-1}\nabla_a\Theta_{,b}-4\Theta^{-2}\Theta_{,a}\Theta_{,b}+g_{ab}(\Theta^{-1}\Box\Theta+\Theta^{-2}|\nabla\Theta|^2),\]
which relates (\ref{ecs4}) to (\ref{es1}) if we take
\bean
\frac{\phi}{1-\phi^2}+\frac{\Theta'}{\Theta}&=&0\nonumber\\
\Theta^2&=&(1-\phi^2)\\
\frac{1}{1-\phi^2}+\frac{\Theta''}{\Theta}+\frac{(\Theta')^2}{\Theta^2}&=&0
\eean
using prime for $d/d\phi$, together with
\[(\tp')^2=\frac32\frac{1}{(1-\phi^2)^2}.\]
These can all be solved with  
\bea
\tp&=&\pm\sqrt\frac32\log\left(\frac{1+\phi}{1-\phi}\right)\label{rel1}\\
\Theta&=&(1-\phi^2)^{1/2},\label{rel2}
\eea
which is one of Bekenstein's transformations.

\section{The effect on total mass}
What effect does this transformation have on the two measures of total mass, at spatial and null infinity?

For the ADM mass of the metric $g$ with source the conformal-scalar field plus $T^{\mathrm{ex}}_{ab}$, we'll assume that there is an asymptotically-flat space-like hypersurface $\mathcal{S}$ 
in the original space-time with metric $g_{ab}$, and that on $\mathcal{S}$
\[\phi=O(1/r),\;\;\phi_{,a}=O(1/r^2).\]
Then on $\mathcal{S}$
\[\Theta=1+O(1/r^2),\;\;\Theta_{,i}=O(1/r^3).\]
The ADM mass can be defined as a limit
\[m=\mbox{const.  }\times\mbox{Lim  }\int(h_{ij,i}-h_{ii,j})dS^j,\]
where the integral is taken over spheres tending to infinity and $h_{ij}$ is the metric on $\mathcal{S}$ induced by $g_{ab}$, in suitable asymptotically Cartesian coordinates. By the same token, 
for the ADM mass of the metric $\tilde{g}$ with source the scalar field plus $\widetilde{T}^{\mathrm{ex}}_{ab}$
\[\tm=\mbox{const.  }\times\mbox{Lim  }\int(\tilde{h}_{ij,i}-\tilde{h}_{ii,j})d\tS^j,\]
but with $\tilde{h}_{ij}=\Theta^2h_{ij}$ and $d\tS^j=\Theta^2 dS^j$. Given the asymptotics of $\Theta$, we conclude $\tm= m$. Thus the ADM mass of the original Einstein-conformal-scalar system, 
possibly with extra sources in $T_{ab}^{\mathrm{ex}}$, is the same as the ADM mass of the transformed system, and this is positive by the standard Positive Mass Theorem since the DEC is satisfied in the 
transformed system. 

For a larger family of transformations of the kind considered here see \cite{fgn}. It seems very likely that the methods used here will show that for these too the ADM mass does not change.

The Bondi mass is more difficult, and the result is different. We'll suppose that $g_{ab}$ is asymptotically-flat (in the sense of weakly-asymptotically-simple) with $\phi=O(1/r)$ at $\scri$. Then 
$\tg_{ab}$ is also asymptotically-flat with $\tp=O(1/r)$ and $\Theta=1$ at $\scri$.

For the definition of the Bondi mass, we'll use the discussion of energy-momentum and angular momentum in \cite{PR}. For any topologically spherical space-like two-surface $S$ one defines a (four-dimensional, complex) 
vector space of \emph{two-surface twistors} as two-component spinor fields satisfying an elliptic equation. Given any two such, say $\omega^A_i,\;i=1,2$, Penrose and Rindler \cite{PR} define a bilinear
\be\label{qlm}
\frac{-i}{4\pi G}\oint\left((\psi_1-\Phi_{01})\omega^0_1\omega^0_2+(\psi_2-\Phi_{11}-\Lambda)(\omega^0_1\omega^1_2+\omega^0_2\omega^1_1)+(\psi_3-\Phi_{21})\omega^1_1\omega^1_2\right)dS.\ee
Here $\psi_i,\Phi_{ij},\Lambda$ are components of curvature, respectively Weyl, trace-free Ricci and scalar. The components of this bilinear are identified with the ten components of momentum and 
angular momentum at $S$. See \cite{PR} for more details, or \cite{t1}, \cite{t2} for some examples of the construction.

Despite a great deal of success in particular situations, there are unresolved difficulties with this definition at a general $S$ in a general space-time. However, the definition can be extended to $\scri$ 
where it recovers the standard definitions of Bondi mass where these have previously been made (for example, for vacuum or Einstein-Maxwell). Therefore we shall use it as our guide with the scalar and 
conformal-scalar, following \cite{BST}. For the Bondi mass, one takes the limit onto $\scri$ of the time-like component of the momentum in (\ref{qlm}) to obtain
\be\label{mb0}
\mbox {Bondi mass }=-c_1\oint\psi_2^0-\Phi_{11}^0-\Lambda^0+\sigma^0\dot{\bar{\sigma}}^0
\ee
where the integration is over a unit sphere, $c_1$ is a positive constant which can be found from the Schwarzschild case, the superscript zero denotes leading nonzero term according to 
the list:
\bean
\psi_2&=&\frac{\psi^0_2}{r^3}+\mbox{  h.o.}\\
\Phi_{11}&=&\frac{\Phi^0_{11}}{r^3}+\mbox{  h.o.}\\
\Lambda&=&\frac{\Lambda^0}{r^3}+\mbox{  h.o.}\\
\sigma&=&\frac{\sigma^0}{r^2}+\mbox{  h.o.}
\eean
writing `h.o.' for higher order terms in $1/r$, and an integration by parts has brought in the terms in $\sigma^0$. The dot indicates derivative with respect to $u$, the parameter along the generators of $\scri$, and 
this simplified expression also relies on $\Phi_{12}=O(r^{-3})$.

For the conformal-scalar case one obtains from (\ref{mb0}) the definition
\be\label{mb1}
m_B=-c_1\oint\psi_2^0-\Phi_{11}^{0\;\mathrm{ex}}+\sigma^0\dot{\bar{\sigma}}^0,
\ee
where $\Phi_{11}^{0\;\mathrm{ex}}$ is a contribution (positive by DEC) from $T_{ab}^{\mathrm{ex}}$. The significant point about this formula 
is that there is no explicit appearance of $\phi$: these Ricci terms, which one might have expected to be present, cancel out.

For the scalar case one finds
\be\label{mb2}
\tm_B=-c_1\oint\tpsi_2^0-\tilde{\Phi}_{11}^{0\;\mathrm{ex}}+\tsigma^0\dot{\bar{\tsigma}}^0+\frac23\tp^0\dot{\tp}^0
\ee
where $\tp^0$ is the leading term in $\phi$, which is assumed to have the asymptotics
\[\tp=\frac{\tp^0}{r}+O(r^{-2}).\]
Given the assumed asymptotics of $\Theta$, one readily discovers that
\[\psi_2^0-\Phi_{11}^{0\;\mathrm{ex}}+\sigma^0\dot{\bar{\sigma}}^0=\tpsi_2^0-\tilde{\Phi}_{11}^{0\;\mathrm{ex}}+\tsigma^0\dot{\bar{\tsigma}}^0,\]
(in fact corresponding terms are equal) so that the two integrands differ only in the term $\frac23\tp^0\dot{\tp}^0$ which comes from the Ricci terms in the tilded space-time. The usual Positive Mass Theorem for the Bondi mass 
will extend to the scalar case, since the DEC is satisfied, so that $\tm_B\geq 0$ but we can't conclude that $m_B\geq 0$ in general. In fact this seems rather unlikely since one would think of $\tp^0$ as the null 
datum in the scalar case and therefore freely disposable on $\scri$. If it is freely disposable then one could choose data to make $m_B<0$, with the definition (\ref{mb1}), 
but now one does not know that this data will lead to a space-time 
sufficiently global for the Positive Mass Theorem to hold (or to avoid having $\phi^2=1$ anywhere in the interior).

As a check on the two definitions of Bondi mass we may calculate the mass-loss, again following \cite{BST}. We find
\[\dot{\tm}_B=-c_1\oint|\dot{\tsigma}^0|^2+\tilde{\Phi}^{0\;\mathrm{ex}}_{22}+2(\dot{\tp}^0)^2\]
which is negative as expected (the term $\tilde{\Phi}^{0\;\mathrm{ex}}_{22}$ is positive by DEC), but
\[\dot{m}_B=-c_1\oint |\dot{\sigma}^0|^2+\Phi^{0\;\mathrm{ex}}_{22}+(\dot{\phi}^0)^2+\phi_0\ddot{\phi}_0\]
which doesn't have a fixed sign because of the $\phi$-terms.
\section{Conclusion}
In conclusion, we've used the transformation of Bekenstein to show that certain asymptotically-flat Einstein-conformal-scalar space-times, possibly with other trace-free sources also present, have positive ADM mass by 
showing that this mass is the same as the ADM mass of the transformed Einstein-scalar space-time. With a natural definition of Bondi mass, the same equality does not hold and there seems no reason in general 
to suppose that the Einstein-conformal-scalar space-time does in fact have positive Bondi mass. That said, one could define the Bondi mass differently for the Einstein-conformal-scalar space-time, namely 
as precisely the Bondi mass of the transformed, Einstein-scalar space-time, whereupon it will be positive. Since this amounts to choosing (\ref{mb2}) but without the tildes in place of (\ref{mb1}) 
as the definition of Bondi mass, it is less artificial than it sounds. 
The question remains open of whether a direct proof of positivity of the ADM mass is possible in 
the cases where the DEC can be locally violated but some average energy-condition still holds, a possibility studied in \cite{few}. 

\section*{Acknowledgement}
I am grateful to Chris Fewster for helpful suggestions, and for drawing my attention to \cite{fgn} and \cite{few}.


\begin{thebibliography}{99}
\bibitem{bek1}
Bekenstein, J 1974 \emph{Exact solutions of Einstein-conformal scalar equations}
Ann.Phys. {\bf{82}} 535--547
%
\bibitem{bek2}
Bekenstein, J 1975 \emph{Black holes with scalar charge} Ann.Phys. {\bf{91}} 75--82
%
\bibitem{BST}
Bi\u{c}\'ak, J, Scholtz, M and Tod, P 2010 \emph{On asymptotically flat solutions of Einstein's equations periodic in time II. Spacetimes with scalar-field sources} Class. Quant. Grav. {\bf 27}, 175011 
%
\bibitem{fgn}
Faraoni, V, Gunzig, E and Nardone, P 1999 \emph{Conformal transformations in classical gravitational theories and in cosmology}
Fund.CosmicPhys. {\bf 20} 121--
%
\bibitem{few}
Fewster, C J and Osterbrink, L W 2006 \emph{Averaged energy inequalities for the nonminimally coupled classical scalar field} Phys. Rev. {\bf D 74}, 044021
%
%
\bibitem{hub}
H\"ubner, P. 1995 \emph{General relativistic scalar-field models and asymptotic flatness} Class.Quant.Grav. {\bf{12}} 791--808
%
%
\bibitem{NP}
Newman, E.T. and Penrose, R. 1968 \emph{New Conservation Laws for Zero Rest-Mass Fields in Asymptotically Flat Space-Time}
Proc. R. Soc. Lond. {\bf{A 305}} 175--204
%
\bibitem{PR}
Penrose, R. and Rindler, W. 1986 \emph{Spinors and space-time} vol II Cambridge Monographs on Mathematical Physics.
 Cambridge University Press, Cambridge.
\bibitem{t1}
Tod, K. P. 1983 \emph{Some examples of Penrose's quasilocal mass construction}  Proc. Roy. Soc. London {\bf A  388} 457--477. 
%
\bibitem{t2}
Tod, K. P. 1990 \emph{Penrose's quasi-local mass} pp 164--188 in \emph{Twistors in mathematics and physics}   London Math. Soc. Lecture Note Ser., 156, Cambridge Univ. Press, Cambridge.
%
\end{thebibliography}
\end{document}